\documentclass[aps,prd]{revtex4}

\usepackage{graphicx}
\usepackage{amsmath}
\usepackage{amssymb}
\usepackage{array}

\def\lg{{\mathchoice{~\raise.58ex\hbox{$<$}\mkern-14.8mu\lower.52ex\hbox{$>$}~}
                    {~\raise.58ex\hbox{$<$}\mkern-14.8mu\lower.52ex\hbox{$>$}~}
                    {\raise.59ex\hbox{{$\scriptscriptstyle <$}}\mkern-12.8mu%
                     \lower.01ex\hbox{{$\scriptscriptstyle >$}}}   {}   }}
\def\gl{{\mathchoice{~\raise.58ex\hbox{$>$}\mkern-12.8mu\lower.52ex\hbox{$<$}~}
                    {~\raise.58ex\hbox{$>$}\mkern-12.8mu\lower.52ex\hbox{$<$}~}
                    {\raise.62ex\hbox{{$\scriptscriptstyle >$}}\mkern-12.0mu%
                     \lower.05ex\hbox{{$\scriptscriptstyle <$}}}  {}    }}

\def\ca{{\mathchoice{~\raise.58ex\hbox{$c$}\mkern-9.0mu\lower.52ex\hbox{$a$}~}
                    {~\raise.58ex\hbox{$c$}\mkern-9.0mu\lower.52ex\hbox{$a$}~}
                    {\raise.59ex\hbox{{$\scriptscriptstyle c$}}\mkern-7.0mu%
		     \lower.01ex\hbox{{$\scriptscriptstyle a$}}}   {} 	}} 
\def\ac{{\mathchoice{~\raise.58ex\hbox{$a$}\mkern-10.0mu\lower.52ex\hbox{$c$}~}
                    {~\raise.58ex\hbox{$a$}\mkern-10.0mu\lower.52ex\hbox{$c$}~}
		    {\raise.62ex\hbox{{$\scriptscriptstyle a$}}\mkern-9.0mu%
		     \lower.05ex\hbox{{$\scriptscriptstyle c$}}}  {} 	}}

\newcommand{\be}{\begin{equation}}
\newcommand{\ee}{\end{equation}}
\newcommand{\ba}{\begin{eqnarray}}
\newcommand{\ea}{\end{eqnarray}}
\newcommand{\ban}{\begin{eqnarray*}}
\newcommand{\ean}{\end{eqnarray*}}
\newcommand \nn {\nonumber}
\newcommand{\sla}{\!\!\!/ \,}

\begin{document}

\title{Universality of Hard-Loop Action}

\author{Alina Czajka}

\affiliation{Institute of Physics, Jan Kochanowski University, Kielce, Poland}

\author{Stanis\l aw Mr\' owczy\' nski}

\affiliation{Institute of Physics, Jan Kochanowski University, Kielce, Poland}
\affiliation{National Centre for Nuclear Research, Warsaw, Poland}

\date{December 17, 2014}

\begin{abstract}

The effective actions of gauge bosons, fermions and scalars, which are obtained within the hard-loop approximation, are shown to have unique forms for a whole class of gauge theories including QED, scalar QED, super QED, pure Yang-Mills, QCD, super Yang-Mills. The universality occurs irrespective of a field content of each theory and of variety of specific interactions. Consequently, the long-wavelength or semiclassical features of plasma systems governed by these theories such as collective excitations are almost identical. An origin of the universality, which holds within the limits of applicability of the hard-loop approach, is discussed.

\end{abstract}



\maketitle

\section{Introduction}

The hard-loop approach is a practical tool to describe plasma systems governed by QED or QCD in a gauge invariant way which is free of infrared divergences, see the reviews \cite{Thoma:1995ju,Blaizot:2001nr,Litim:2001db,Kraemmer:2003gd}. Initially the approach was developed within the thermal field theory \cite{Braaten:1989mz,Braaten:1990az} but it was soon realized that it can be formulated in terms of quasiclassical kinetic theory \cite{Blaizot:1993be,Kelly:1994dh}. The plasma systems under consideration were assumed to be in thermodynamical equilibrium but the methods can be naturally generalized to plasmas out of equilibrium \cite{Pisarski:1997cp,Mrowczynski:2000ed}. 

An elegant and concise formulation of the hard-loop approach is achieved by introducing an effective action derived for equilibrium and non-equilibrium systems in \cite{Taylor:1990ia,Frenkel:1991ts,Braaten:1991gm} and \cite{Pisarski:1997cp,Mrowczynski:2004kv}, respectively. The action is a key quantity that encodes an infinite set of hard-loop $n$-point functions. A whole gamut of long-wavelength characteristics of a plasma system is carried by the functions. In particular, the two-point functions or self-energies provide response functions like permeabilities or susceptibilities which control various screening lengths. The self-energies also determine a spectrum of collective excitations (quasiparticles) that is a fundamental characteristic of any many-body system. 

One wonders how much a given plasma characteristic is different for different plasma systems. It has been known for a long time that the self-energies of gauge bosons in the long-wavelength limit are of the same structure for QED and QCD plasmas \cite{Weldon:1982aq}. Consequently, the collective excitations and many other characteristics are the same, or almost the same, in the two plasma systems \cite{Mrowczynski:2007hb}. However, it should be remembered that these systems are so similar in the domain of validity of the hard-loop approach that is when the momentum scale of collective degrees of freedom is neither too long nor too short. We return to this problem at the end of Sec.~\ref{sec-self-en}.

Comparing systematically supersymmetric plasmas to their non-supersymmetric counterparts, we have considered \cite{Czajka:2010zh,Czajka:2011zn,Czajka:2012gq} a whole class of gauge theories including Abelian cases: QED, scalar QED, and ${\cal N}\!=\!1$ super QED and nonAbelian ones: pure Yang-Mills, QCD, and ${\cal N}\!=\!4$ super Yang-Mills. We have observed that the self-energies of gauge bosons, fermions and scalars, which are computed in the hard-loop approximation, have unique structures for all considered theories irrespective of a field content and of variety of specific interactions. Consequently, the hard-loop effective actions are essentially the same and so are long-wavelength characteristics of plasma systems governed by the gauge theories of interest. Although our findings are partially presented in \cite{Czajka:2010zh,Czajka:2011zn,Czajka:2012gq}, we have decided to collect all our results in this paper and to systematically elaborate on the problem. We explain an origin of the universality, that is, how it happens that the microscopically different systems are very similar to each other in the long-wavelength limit. Physical consequences of the universality and its limitations are also discussed. 

Our paper is organized as follows. In the next section, we briefly present the gauge theories taken into consideration. The differences and similarities of the theories are underlined. Sec.~\ref{sec-self-en} is devoted to the self-energies of gauge bosons, fermions and scalars which are computed in the hard-loop approximation. Validity of the approximation is also explained here. Knowing the self-energies, the effective action of the hard-loop approach is derived in Sec.~\ref{sec-eff-act}. An origin of the universality of the hard-loop action, its physical consequences and limitations are discussed in Sec.~\ref{sec-discussion} which concludes our study.

Throughout the paper we use the natural system of units with $c= \hbar = k_B =1$; our choice of the signature of the metric tensor is $(+ - - -)$.

\section{Gauge theories under consideration}
\label{sec-theories}

We briefly present here the gauge theories under consideration stressing differences and similarities among them. We start with 
QED of the commonly known Lagrangian density that is 
\ba
\label{L-QED}
{\cal L}_{\textrm{QED}} &=& -\frac{1}{4}F^{\mu \nu} F_{\mu \nu} 
+ i\bar \Psi D\!\sla \Psi,
\ea
where the strength tensor $F^{\mu \nu}$ is expressed through the electromagnetic four-potential $A^\mu$ as $F^{\mu \nu} \equiv \partial^\mu A^\nu - \partial^\nu A^\mu$,  $\Psi$ is the Dirac bispinor electron field, $ D\!\sla \equiv \gamma^\mu D_\mu$ and the covariant derivative equals $D^\mu \equiv \partial^\mu - ie A^\mu$. Since we are interested in ultrarelativistic plasmas, where the plasma constituents are treated as massless, the mass term is neglected in Eq.~(\ref{L-QED}) and in all other cases under study. As well known, the Lagrangian (\ref{L-QED}) describes a system of electrons, positrons and photons governed by a long-range electromagnetic interaction represented by the term $e\bar\Psi \gamma^\mu \Psi A_\mu$.

Replacing the electron bispinor $\Psi$ with the scalar complex field $\Phi$, we get the scalar electrodynamics of spinless charges and the Lagrangian reads
\ba
\label{L-ScQED}
{\cal L}_{\textrm{scalar\,QED}} &=& -\frac{1}{4}F^{\mu \nu} F_{\mu \nu} 
- (D^\mu \Phi)^* D_\mu \Phi.
\ea
Except for the interaction terms $e(\partial^\mu \Phi^*)  \Phi  A_\mu$ and $ e\Phi^* (\partial^\mu \Phi)  A_\mu$, there is a four-boson coupling $e^2\Phi^* \Phi A^\mu A_\mu$. Such a contact interaction is qualitatively different than that caused by a massless particle exchange. In absence of other interactions, it gives the scattering which is isotropic in the center-of-mass frame of colliding particles with characteristic energy and momentum transfers which are much bigger than those in one photon-exchange processes. 

A peculiar combination of QED and scalar QED is ${\cal N}\!=\!1$ super QED, see {\it e.g.} \cite{Binoth:2002xg}, with the Lagrangian of the form 
\ba
\label{L-SUSY-QED}
{\cal L}_{\textrm{super\,QED}} &=&{\cal L}_{\textrm{QED}}
+\frac{i}{2} \bar \Lambda \partial \sla \Lambda
+(D_\mu \Phi_L)^*(D^\mu \Phi_L) + (D_\mu^* \Phi_R)(D^\mu \Phi_R^*)
\\ \nn
&& +\sqrt{2} e \big( \bar \Psi P_R \Lambda \Phi_L - \bar \Psi P_L \Lambda \Phi_R^*
+ \Phi_L^* \bar \Lambda P_L \Psi - \Phi_R \bar \Lambda P_R \Psi \big)
- \frac{e^2}{2} \big( \Phi_L^* \Phi_L - \Phi_R^* \Phi_R \big)^2 ,
\ea
where $\Lambda$ is the Majorana bispinor photino field, $\Phi_L$ and $\Phi_R$ represent the scalar left and right selectrons; the projectors $P_L$ and $P_R$ are defined in a standard way $P_L \equiv \frac{1}{2}(1 - \gamma_5)$ and $P_R \equiv \frac{1}{2}(1 + \gamma_5)$. The supersymmetric extension of QED describes a mixture of photons, Majorana and Dirac fermions, and scalars of two types with a variety of interactions. Except for the long-range one-photon exchanges, we have four-boson couplings and the Yukawa interactions of non-electromagnetic nature. The complete list of elementary processes, which is given in \cite{Czajka:2011zn}, is thus very long and it makes the supersymmetric plasma very different  at the microscopic level from the usual electromagnetic ones.

The first nonAbelian plasma under study is that governed by the pure Yang-Mills theory with the ${\rm SU}(N_c)$ gauge group.  The Lagrangian of gluodynamics is
\ba
\label{L-YM}
{\cal L}_{\textrm{YM}} =-\frac{1}{4}F^{\mu \nu}_a F_{\mu \nu}^a,
\ea
where $a,b= 1,2, \dots N_c^2-1$ and the chromodynamic strength tensor $F_a^{\mu \nu}$ is expressed by the four-potential $A_a^\mu$ as $F^{\mu \nu}_a \equiv \partial^\mu A^\nu_a - \partial^\nu A^\mu_a + g f^{abc} A^\mu_b A^\nu_c$ with $g$ being the coupling constant and $ f^{abc}$ the structure constant of  the ${\rm SU}(N_c)$ group. Due to the self-interaction of Yang-Mills fields, there is the three- and four-gluon coupling. 

Enriching the pure gluodynamics with (massless) quarks of $N_f$ flavors, which belong to the fundamental representation of the ${\rm SU}(N_c)$ gauge group, we get QCD with the Lagrangian 
\ba
\label{L-QCD}
{\cal L}_{\textrm{QCD}} = {\cal L}_{\textrm{YM}} +i \bar \Psi_i D\!\sla \Psi_i,
\ea
where $i= 1,2,\dots N_f$ and the covariant derivative equals $D_\mu \equiv \partial_\mu  - ig \tau^a A^a_\mu $ with $\tau^a$ being the generator of fundamental representation of the ${\rm SU}(N_c)$ group. Except for the three- and four-gluon couplings, gluons also interact with the color quark current. 

Finally, the Lagrangian of ${\mathcal N} \! = \! 4$ super Yang-Mills theory, see  {\it e.g.} \cite{Yamada:2006rx}, can be written as
\ba
\label{L-Super-YM}
{\cal L}_{\textrm{super\,YM}} &=&  {\cal L}_{\textrm{YM}}
+\frac{i}{2}\bar \Psi_i^a (D\!\sla \Psi_i)^a
+\frac{1}{2}(D_\mu \Phi_A)_a (D^\mu \Phi_A)_a
\\ [2mm] \nn
&&
-\frac{1}{4} g^2f^{abe} f^{cde} \Phi_A^a \Phi_B^b \Phi_A^c \Phi_B^d
-i\frac{g}{2} f^{abc} \Big( \bar \Psi_i^a  \alpha_{ij}^p  X_p^b \Psi_j^c  
+i\bar \Psi_i^a \beta_{ij}^p\gamma_5  Y_p^b \Psi_j^c \Big),
\ea
where instead of quarks we have four Majorana fermions represented by $\Psi_i^a$ with $i, j = 1,2,3,4$ and six real scalar fields which are assembled in the multiplet $\Phi = (X_1, Y_1, X_2, Y_2, X_3, Y_3)$. The components of $\Phi$ are either denoted as $X_p$ for scalars, and $Y_p$ for pseudoscalars, with $p,q =1,2,3$ or as $\Phi_A$ with $A, B=1,2, \dots 6$. The $4 \times 4$ matrices $\alpha^p, \beta^p$ satisfy the relations
\be
\label{alpha-beta-relations}
\{\alpha^p, \alpha^q \} = - 2 \delta^{p q},
\;\;\;\;\;\;\;
\{\beta^p, \beta^q \} = - 2 \delta^{p q},
\;\;\;\;\;\;\;
[ \alpha^p, \beta^q] = 0 .
\ee
In the super Yang-Mills theory all fields belong to the adjoint representation of the ${\rm SU}(N_c)$ gauge group and the covariant derivative is $D_\mu^{ab} \equiv \partial_\mu \delta^{ab}  + g  f^{abc} A^c_\mu $. As in QCD there are the three- and four-gluon couplings and the gluon interaction with the color fermion current. Additionally there are the four-boson couplings $g^2\Phi_A \Phi_A A^\mu A_\mu$ and $g^2\Phi_A \Phi_B \Phi_A \Phi_B$. There is also the Yukawa interaction of fermions with scalars. The complete list of elementary interactions, which is given in \cite{Huot:2006ys}, is again rather long and it makes the super Yang-Mills plasma quite different at the microscopic level from the gluodynamic or QCD plasmas.

\section{Self-energies}
\label{sec-self-en}

Our objective is to derive the effective action of all considered theories in the hard-loop approximation. The action $S$ can be found {\it via} the respective self-energies which are the second functional derivatives of $S$ with respect to the given fields. Thus, the self-energies of gauge boson, fermion and scalar fields equal
\ba
\label{se-Pi}
\Pi^{\mu \nu}(x,y) &=& \frac{\delta^2 S}{\delta A_\mu(x) \,\delta A_\nu(y)}, 
\\ [2mm]
\label{se-Sigma}
\Sigma (x,y) &=& \frac{\delta^2 S}{\delta \bar\Psi (x) \,\delta \Psi(y)}, 
\\ [2mm]
\label{se-P}
P(x,y) &=& \frac{\delta^2 S}{\delta \Phi^*(x) \,\delta \Phi (y)},
\ea
where the field indices, which are different for different theories under consideration, are suppressed. The action will be obtained in the subsequent section by integrating the formulas (\ref{se-Pi})-(\ref{se-P}) over the respective fields. 

We compute the self-energies, which enter Eqs.~(\ref{se-Pi})-(\ref{se-P}), diagrammatically. The plasma systems under study are assumed to be homogeneous in coordinate space (translationally invariant), locally colorless and unpolarized, but the momentum distribution may be arbitrary. Therefore, we use the Keldysh-Schwinger or real-time formalism, explained in {\it e.g.} \cite{Mrowczynski:1992hq}, which allows one to describe many-body systems both in and out of equilibrium. 

In the Tables \ref{tab-gluons}, \ref{tab-fermions}, and \ref{tab-scalars} we present the diagrams of the lowest order (one loop)  contributions to the self-energies of gauge bosons, fermions, and scalars, respectively, for all studied theories. Needless to say that the coupling constant $g$ (or $e$) is assumed to be small. Since the Feynman gauge is used, the ghost loop contributes to the gluon polarization tensor. The curly, plain, dotted, and dashed lines denote, respectively, the gauge, fermion, ghost, and scalar fields.

\begin{table}[t]
\caption{\label{tab-gluons} The diagrams of the lowest order contributions to the polarization tensors.}
\begin{tabular}{m{5cm} m{8cm}}
\hline
\vspace{1mm}
Plasma system & Lowest order diagrams
\\
\hline 
\\
QED & \includegraphics[scale=0.2]{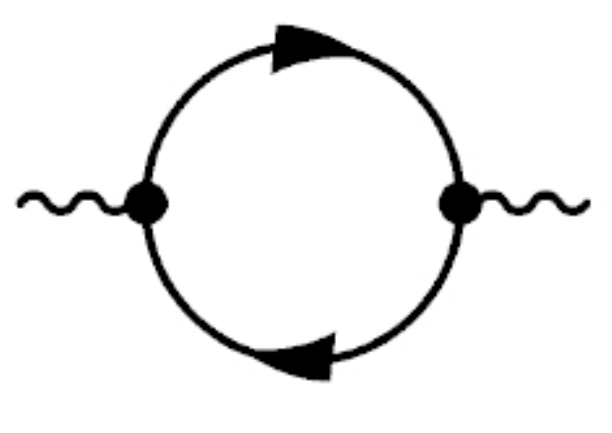}
\\
scalar QED & \includegraphics[scale=0.2]{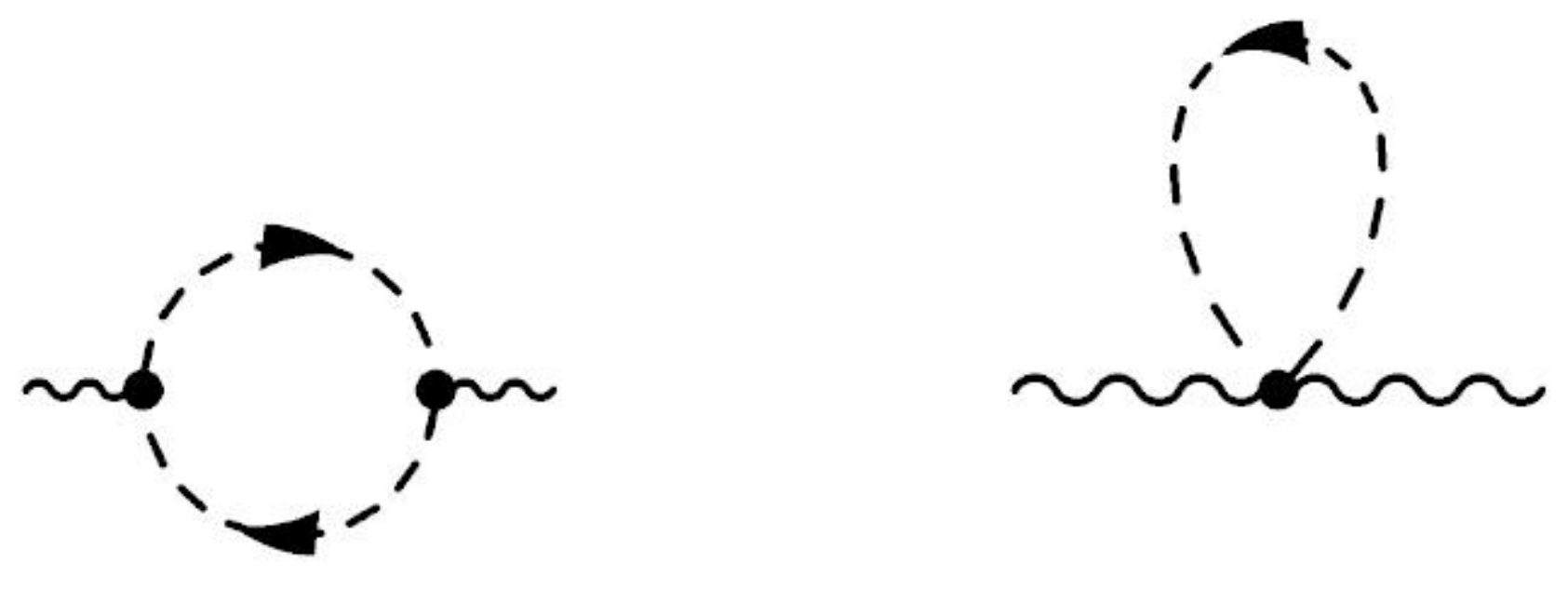}
\\
${\cal N}=1$ super QED & \includegraphics[scale=0.2]{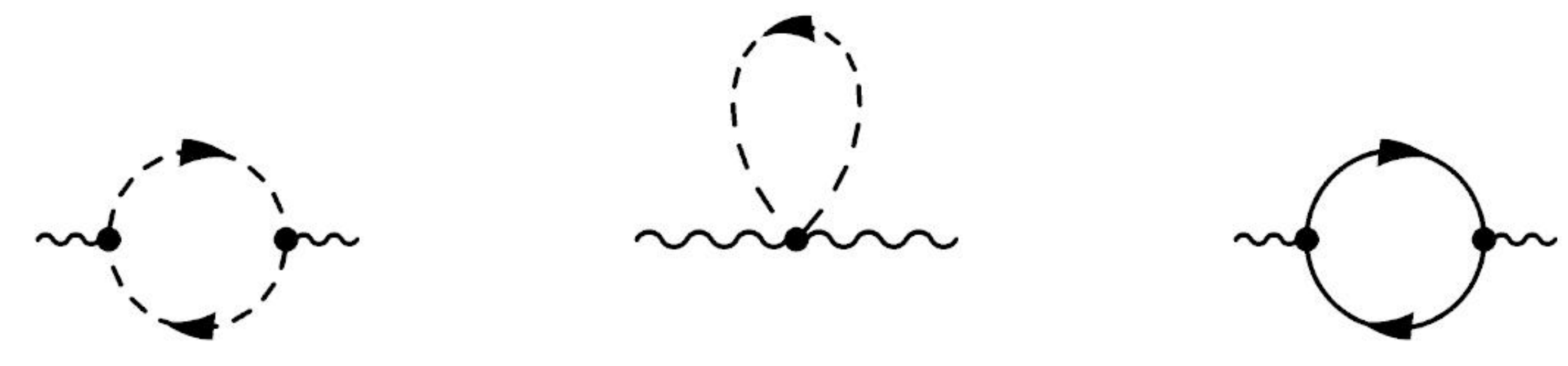}
\\
Yang-Mills & \includegraphics[scale=0.2]{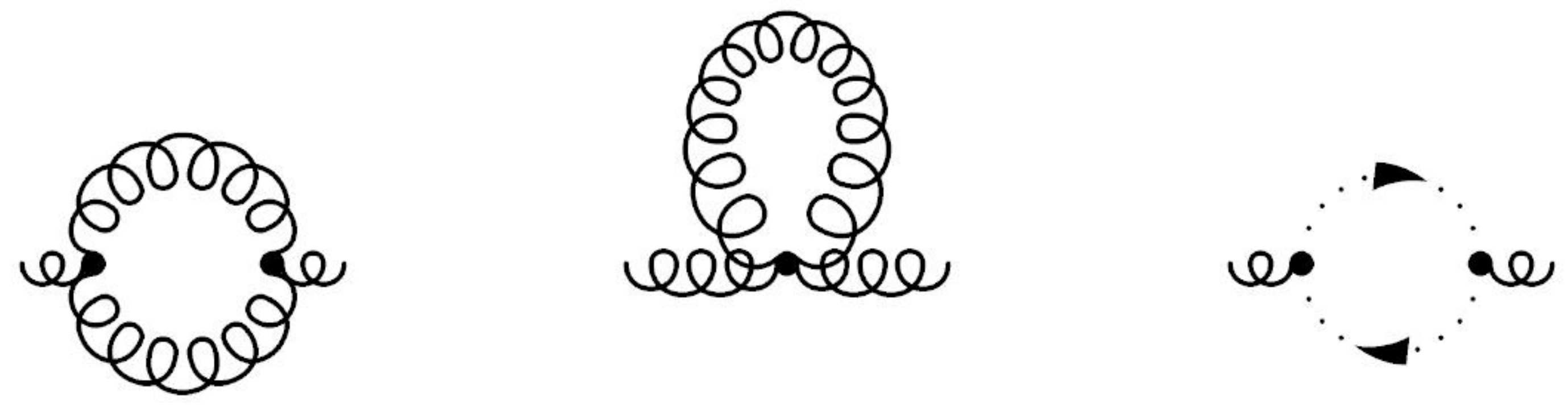}
\\
QCD & \includegraphics[scale=0.2]{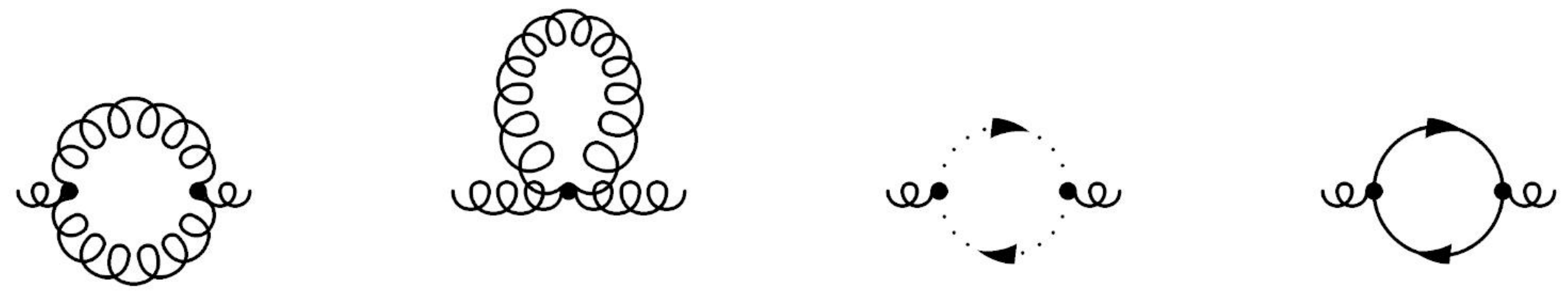}
\\
${\cal N}=4$ super Yang-Mills & \includegraphics[scale=0.2]{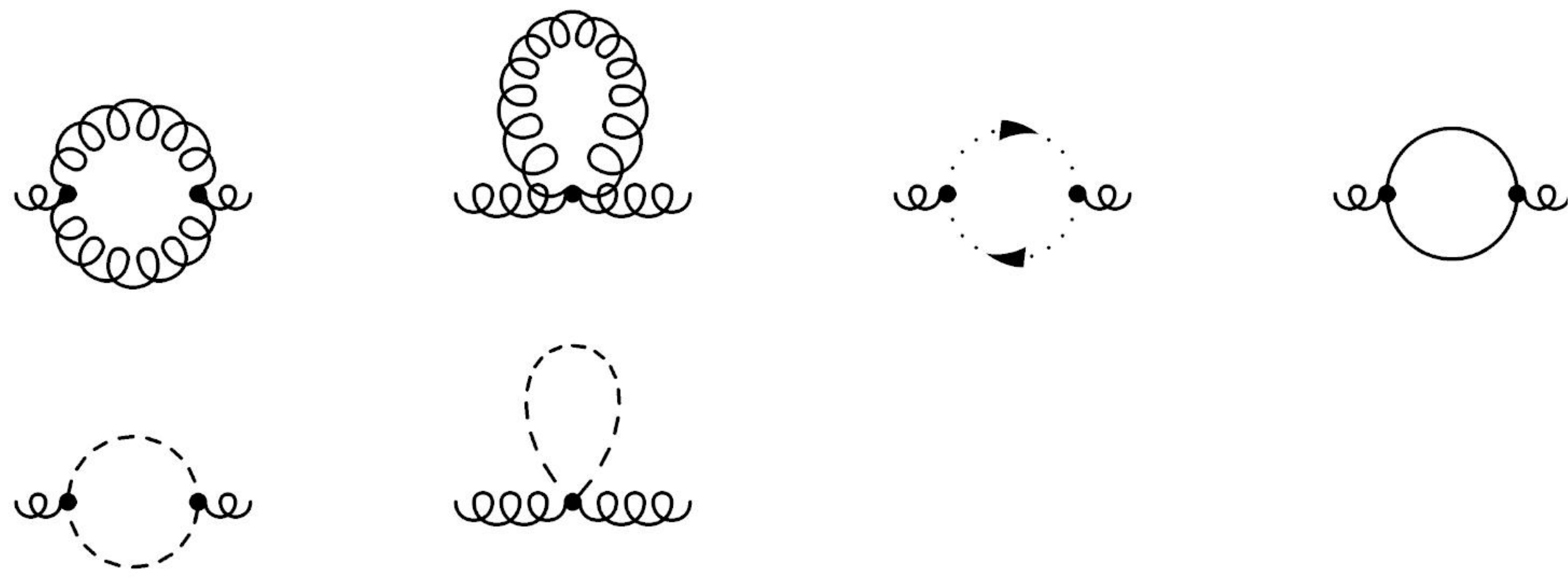}
\\
\hline
\end{tabular}
\end{table}

\begin{table}[b]
\caption{\label{tab-factors-gluons} The factors entering the polarization tensors.}
\begin{tabular}{lcc}
\hline
Plasma system & $C_{\Pi}$ &  $f_{\Pi}({\bf p})$
\\
\hline 
\\
\vspace{1mm}
QED & $e^2$ &  $2 f_e({\bf p}) + 2 \bar f_e({\bf p})$
\\
\vspace{1mm}
scalar QED & $e^2$ &  $f_s({\bf p}) + \bar f_s({\bf p})$
\\
\vspace{1mm}
${\cal N}=1$ super QED & $e^2$ &  $2 f_e({\bf p}) + 2 \bar f_e({\bf p}) + 2f_s({\bf p}) + 2\bar f_s({\bf p}) $
\\
\vspace{1mm}
Yang-Mills & $~~~g^2 N_c  \delta^{ab}~~~$ & $2 f_g ({\bf p})$
\\
\vspace{1mm}
QCD &  $g^2 N_c  \delta^{ab}$ & $2 f_g ({\bf p}) + \frac{N_f}{N_c} \big(f_q ({\bf p}) + \bar f_q ({\bf p}) \big) $
\\
\vspace{1mm}
${\cal N}=4$ super Yang-Mills &  $g^2 N_c  \delta^{ab}$ & $2f_g ({\bf p}) + 8f_f ({\bf p}) + 6f_s ({\bf p}) $
\\
\hline
\end{tabular}
\end{table}

As seen in Table \ref{tab-gluons}, both the number of diagrams contributing to the polarization tensor and their forms are different for each theory. We have the fermion, scalar and gluon loops and the scalar and gluon tadpoles which differently depend on the external momentum. Accordingly, there is no surprise that the polarization tensors $\Pi^{\mu \nu}(k)$ are quite different for each theory. However, when the external momentum $k$ is much smaller than the internal momentum $p$, which flows along the loop and is carried by a plasma constituent, that is when the hard-loop approximation $(k \ll p)$ is applied, we get a very striking result: the (retarded) polarization tensors of all theories are of the same form 
\be
\label{Pi-k-final}
\Pi^{\mu \nu}(k)
= C_{\Pi}
\int \frac{d^3p}{(2\pi)^3}
\frac{f_{\Pi}({\bf p})}{E_p} 
\frac{k^2 p^\mu p^\nu - (k^\mu p^\nu + p^\mu k^\nu - g^{\mu \nu} (k\cdot p))
(k\cdot p)}{(k\cdot p + i 0^+)^2},
\ee
where $C_{\Pi}$ is the factor and $f_{\Pi}({\bf p})$ the effective distribution function of plasma constituents which are both given in Table \ref{tab-factors-gluons} for each plasma system. $f_e({\bf p})$ and $\bar f_e({\bf p})$ denote the electron and, respectively, positron distribution functions. The meaning of other functions can be easily guessed. We only add that $f_{\tilde\gamma} ({\bf p})$ is the distribution function of photinos. All functions are normalized in such a way that
\be
\rho_f = \int \frac{d^3p}{(2\pi)^3} f_f ({\bf p})
\ee
is density of particles $f$ of a given spin and color, if any. Particles of the same type but different spin and/or color are assumed to have the same momentum distribution. The left and right selectrons in ${\cal N}=1$ super QED have the same  momentum distribution as well. It is also assumed that quarks of all flavors, similarly as all fermions and all scalars in $\mathcal{N}=4$ super Yang-Mills plasma, have the same momentum distribution. In case of non-supersymmetric plasmas, there is subtracted from the formula (\ref{Pi-k-final}) the (infinite) vacuum contribution which otherwise survives when $f_{\Pi}({\bf p})$ is sent to zero. The subtraction is not needed for the supersymmetric theories where the vacuum effect cancels out.  The polarization tensor (\ref{Pi-k-final}), which is chosen to obey the retarded initial condition, is symmetric in Lorentz indices, $\Pi^{\mu\nu}(k) = \Pi^{\nu\mu}(k)$, and transverse, $k_\mu \Pi^{\mu\nu}(k)=0$, and thus it is gauge independent. We note that the transversality of $\Pi^{\mu\nu}(k)$ is {\em not} an assumption but it automatically results from the calculations, the details of which are given in   \cite{Czajka:2010zh,Czajka:2012gq,Czajka:2014eha} for the electromagnetic theories, $\mathcal{N}=4$ super Yang-Mills, and QCD, respectively. In case of nonAbelian theories, the transversality of $\Pi^{\mu\nu}(k)$ requires to include the Faddeev-Popov ghosts when the calculations are performed in a covariant gauge. The problem how to include the ghosts in the Keldysh-Schwinger formalism is discussed in \cite{Czajka:2014eha}. 

\begin{table}[t]
\caption{\label{tab-fermions} The diagrams of the lowest order contributions to the fermion self-energies.}
\begin{tabular}{m{5cm} m{8cm}}
\hline
Plasma system & Lowest order diagrams
\\
\hline
\\
QED & \includegraphics[scale=0.2]{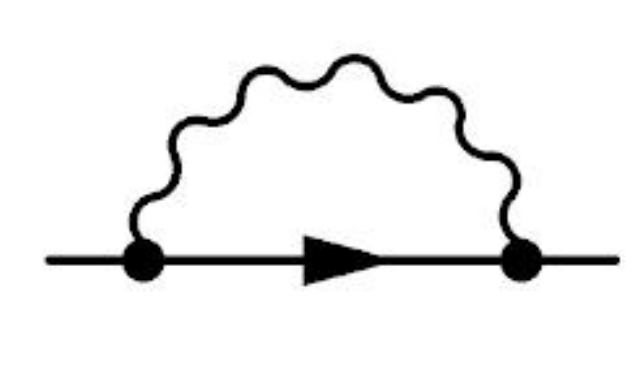}
\\
electron in ${\cal N}=1$ super QED & \includegraphics[scale=0.2]{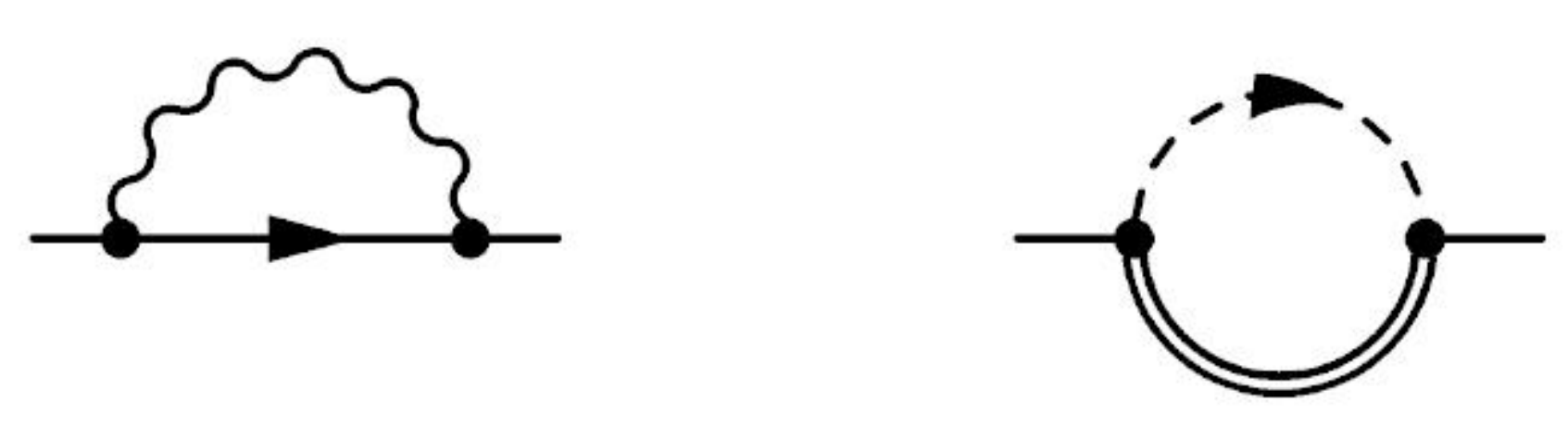}
\\
photino in ${\cal N}=1$ super QED & \includegraphics[scale=0.2]{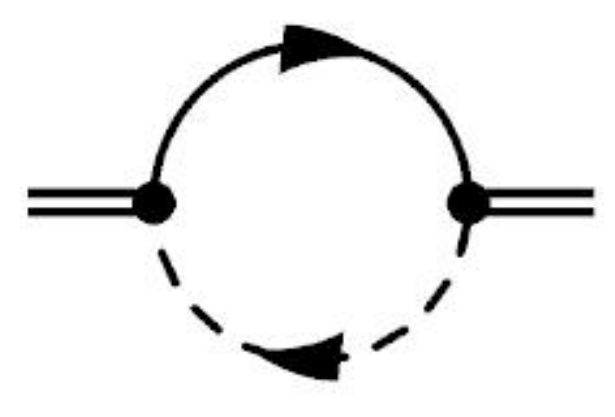}
\\
QCD & \includegraphics[scale=0.2]{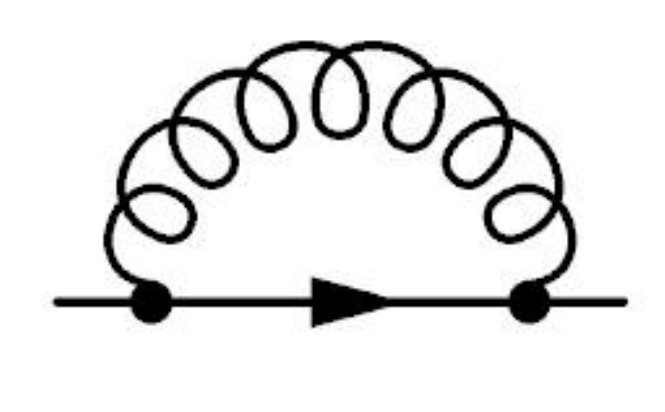}
\\
${\cal N}=4$ super Yang-Mills & \includegraphics[scale=0.2]{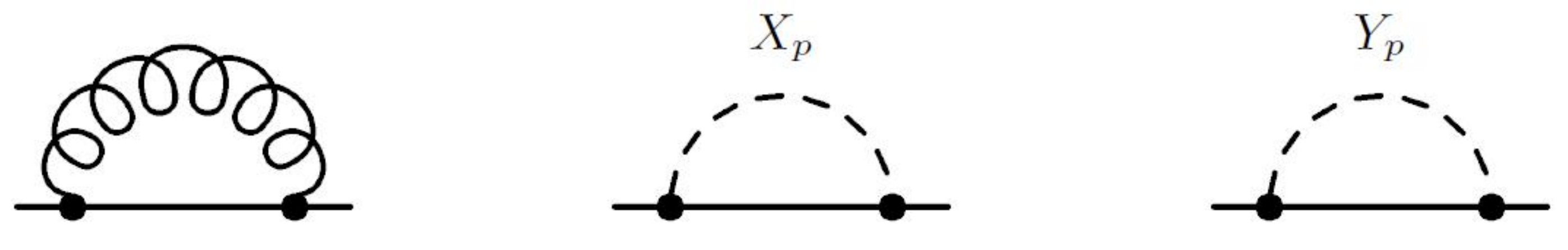}
\\
\hline
\end{tabular}
\end{table}

One wonders how the universality of the polarization tensor emerges. This is not the case that every one-loop contribution behaves in the same way in the long-wavelength limit. Just the opposite, the fermion loops contribute differently than boson ones, and the tadpoles are different than the loops. However, every subset of diagrams which is, as a sum of the diagrams, gauge independent, has the same  long-wavelength limit. For example, in the $\mathcal{N}=4$ super Yang-Mills theory we have three such subsets. The first one is simply the fermion loop, the second one is the sum of the scalar loop and scalar tadpole, and the third gauge independent subset is the sum of the gluon loop, the gluon tadpole and the ghost loop. We also note that the universality holds within the domain of validity of the hard-loop approximation which is explained at the end of this section after all self-energies of interest are given. A physical origin of the universality is discussed in Sec.~\ref{sec-discussion}.

\begin{table}[b]
\caption{\label{tab-factors-fermions} The factors entering the fermion self-energies.}
\begin{tabular}{lcc}
\hline
Plasma system & $C_{\Sigma}$ &  $f_{\Sigma}({\bf p})$
\\
\hline
\\
\vspace{1mm}
QED &  $\frac{e^2}{2}$ &  $ 2 f_\gamma ({\bf p}) + f_e({\bf p}) + \bar f_e({\bf p}) $
\\
\vspace{1mm}
electron in ${\cal N}=1$ super QED &  $\frac{e^2}{2}$ &  $ 2 f_\gamma ({\bf p}) + f_e({\bf p}) + \bar f_e({\bf p}) + 2 f_{\tilde\gamma} ({\bf p}) + f_s({\bf p}) + \bar f_s({\bf p}) $
\\
\vspace{1mm}
photino in ${\cal N}=1$ super QED &  $\frac{e^2}{2}$ &  $f_e({\bf p}) + \bar f_e({\bf p}) + f_s({\bf p}) + \bar f_s({\bf p}) $
\\
\vspace{1mm}
QCD &  $~~~ \frac{g^2}{2} \frac{N_c^2-1}{2 N_c} \delta^{m n} \delta^{ij}~~~$ & $2 f_g ({\bf p}) + N_f \big(f_q ({\bf p}) + \bar f_q ({\bf p}) \big) $
\\
\vspace{1mm}
${\cal N}=4$ super Yang-Mills & $\frac{g^2}{2} N_c  \delta^{ab} \delta^{ij}$ & $2f_g ({\bf p}) + 8f_f ({\bf p}) + 6f_s ({\bf p}) $
\\
\hline
\end{tabular}
\end{table}

In Table \ref{tab-fermions} there are listed the lowest order contributions to the fermion self-energies of every theory. In case of the  ${\cal N}=1$ super QED, there are the Dirac fermions and Majorana fermions which have to be treated differently. As in case of the polarization tensor, the fermion self-energies $\Sigma(k)$ are quite different for each theory. However, when the external momentum $k$ is much smaller than the internal momentum $p$ that is when the hard-loop approximation is applied, the (retarded) self-energies of all theories are of the same form 
\ba
\label{Si-k-final}
\Sigma(k) &=& C_{\Sigma}
\int \frac{d^3p}{(2\pi )^3}
\frac{f_{\Sigma}({\bf p})}{E_p}  \, \frac{p\sla}{k\cdot p + i 0^+},
\ea
where $C_{\Sigma}$ and $f_{\Sigma}({\bf p})$ are both given in Table \ref{tab-factors-fermions} for each plasma system. The indices $m, n = 1, 2, \dots N_c$ label quark colors in the fundamental representation of ${\rm SU}(N_c)$ group.

Table \ref{tab-scalars} shows the diagrams of the lowest order contributions to the scalar self-energy of three theories where scalars occur.  As in case of the polarization tensors and fermion self-energies, the self-energy of scalars $P(k)$ are quite different for each theory. However, within the hard-loop approximation we obtain the amazingly repetitive result - the scalar self-energies of all theories have the same form 
\be
\label{P-k-final}
P(k) = - C_P \int \frac{d^3p}{(2\pi)^3} \frac{f_P({\bf p})}{E_p},
\ee
where $C_P$ and $f_P({\bf p})$ are both given in Table \ref{tab-factors-scalars} for each plasma system. As seen, the self-energy (\ref{P-k-final}) is real, negative and it is independent of the wave vector $k$.

\begin{table}[t]
\caption{\label{tab-scalars} The diagrams of the lowest order contributions to the scalar self-energies.}
\begin{tabular}{m{5cm} m{8cm}}
\hline
Plasma system & Lowest order diagrams
\\
\hline
\\
scalar QED &  \includegraphics[scale=0.2]{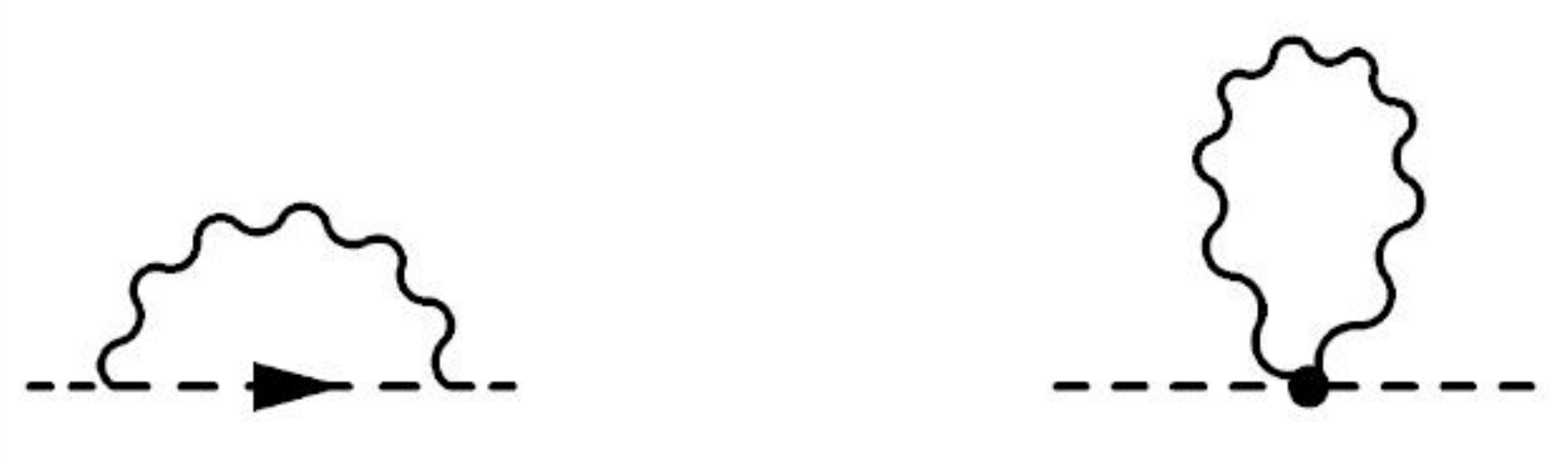}
\\
${\cal N}=1$ super QED & \includegraphics[scale=0.2]{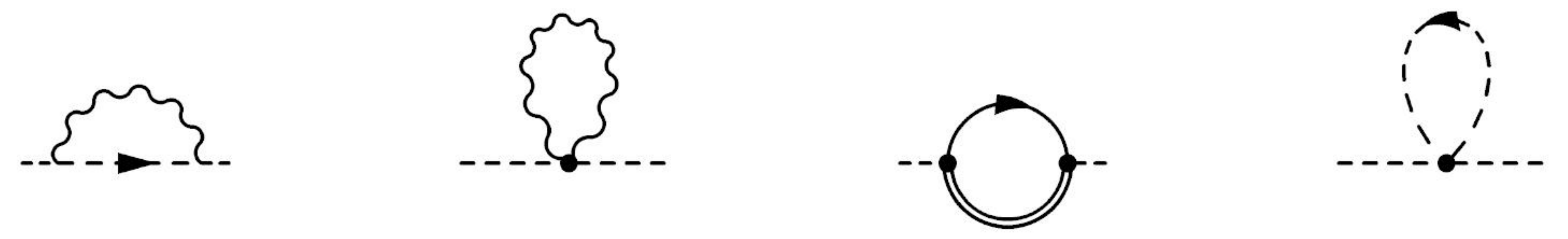}
\\
${\cal N}=4$ super Yang-Mills & \includegraphics[scale=0.2]{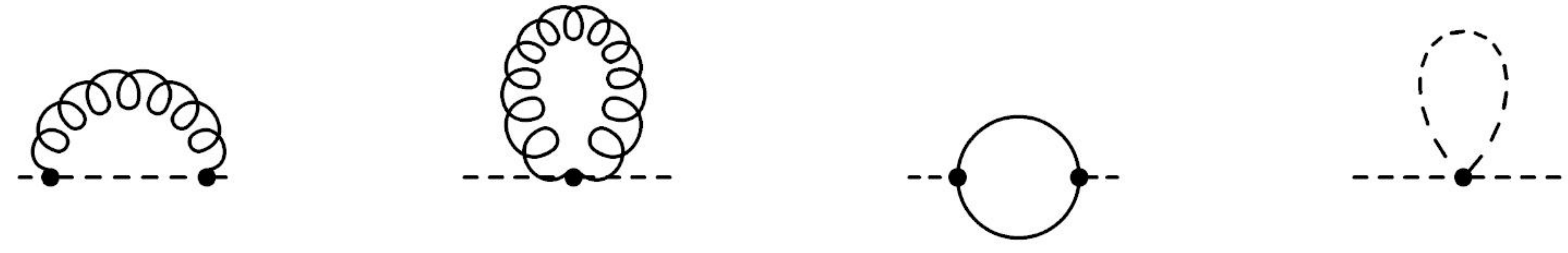}
\\
\hline
\end{tabular}
\end{table}
   
The universal expressions of the self-energies (\ref{Pi-k-final}), (\ref{Si-k-final}), and (\ref{P-k-final}) have been obtained in the hard-loop approximation that is when the external momentum $k$ is much smaller than the internal momentum $p$ which is carried by a plasma constituent. However, it appears that the self-energies (\ref{Pi-k-final}), (\ref{Si-k-final}), and (\ref{P-k-final}) are valid when the external momentum $k$ is not too small. It is most easily seen in case of the fermion self-energy (\ref{Si-k-final}) which diverges as $k \rightarrow 0$. When we deal with an equilibrium (isotropic) plasma of the temperature $T$, the characteristic momentum of (massless) plasma constituents is of the order $T$. One observes that if the external momentum $k$ is of the order $g^2 T$, which is the so-called {\it magnetic} or  {\it ultrasoft} scale, the self-energy (\ref{Si-k-final}) is not perturbatively small as it is of the order ${\cal O}(g^0)$. Therefore, the expression (\ref{Si-k-final}) is meaningless for $k \le g^2 T$.  Since $k$ must be much smaller than $p \sim T$, one arrives to the well-known conclusion that the self-energy (\ref{Si-k-final}) is valid at the {\it soft} scale that is when $k$ is of the order $gT$.  Analyzing higher order corrections to the self-energies (\ref{Pi-k-final}), (\ref{Si-k-final}), (\ref{P-k-final}), one shows that they are indeed valid for $k \sim gT$ and they break down at the magnetic scale because of the infrared problem of gauge theories, see {\it e.g.} \cite{Lebedev:1989ev}
or the review \cite{Kraemmer:2003gd}. When the momentum distribution of plasma particles is anisotropic, instead of the temperature $T$, we have a characteristic four-momentum ${\cal P}^\mu$ of plasma constituents and the hard-loop approximation requires that ${\cal P}^\mu \gg k^\mu$ which should be understood as a set of four conditions for each component of the four-momentum $k^\mu$. Validity of the self-energies (\ref{Pi-k-final}), (\ref{Si-k-final}), and (\ref{P-k-final}) is then limited to $k^\mu \sim g {\cal P}^\mu$.  

\begin{table}[b]
\caption{\label{tab-factors-scalars} The factors entering the scalar self-energies.}
\begin{tabular}{lcc}
\hline
Plasma system &  $C_P$ &  $f_P({\bf p})$
\\
\hline
\\
\vspace{1mm}
scalar QED & $e^2$ & $ 2 f_\gamma ({\bf p}) + f_s({\bf p}) + \bar f_s({\bf p}) $
\\
\vspace{1mm}
${\cal N}=1$ super QED & $e^2$ & $ 2 f_\gamma ({\bf p}) + f_e({\bf p}) + \bar f_e({\bf p}) + 2 f_{\tilde\gamma} ({\bf p}) + f_s({\bf p}) + \bar f_s({\bf p}) $
\\
\vspace{1mm}
${\cal N}=4$ super Yang-Mills & $~~~ g^2 N_c  \delta^{ab} \delta^{AB}~~~$ & $2f_g ({\bf p}) + 8f_f ({\bf p}) + 6f_s ({\bf p}) $
\\
\hline
\end{tabular}
\end{table}

\section{Effective action}
\label{sec-eff-act}

Having the self-energies $\Pi^{\mu\nu}(k),\; \Sigma(k)$, and $P(k)$ given by Eqs.~(\ref{Pi-k-final}), (\ref{Si-k-final}), and (\ref{P-k-final}), respectively, we can reconstruct the effective action. Integrating the formulas (\ref{se-Pi})-(\ref{se-P}) over the respective fields, we obtain the Lagrangian densities
\ba
\label{action-A-1}
{\cal L}^{A}_2(x) &=&
\frac{1}{2} \int d^4y \; A_\mu(x) \Pi^{\mu \nu}(x-y) A_\nu(y) ,
\\ [2mm]
\label{action-Psi-1}
{\cal L}^{\Psi}_2(x) &=&
\int d^4y \; \bar\Psi(x) \Sigma (x-y) \Psi(y) ,
\\ [2mm]
\label{action-Phi-1}
{\cal L}^{\Phi}_2(x) &=&
\int d^4y \; \Phi^*(x) P(x-y) \Phi(y) .
\ea
In case of ${\cal N}=4$ super Yang-Mills, where the scalar fields are real, there is an extra factor 1/2 in the r.h.s of Eq.~(\ref{action-Phi-1}). The subscript `2' indicates that the above effective actions generate only two-point functions. We omit the field indices in Eqs.~(\ref{action-A-1})-(\ref{action-Phi-1}) to keep the expressions applicable to all considered theories. The action is obviously related to the Lagrangian density as $S = \int d^4x \, {\cal L}$. Using the explicit expressions of  the self-energies (\ref{Pi-k-final}), (\ref{Si-k-final}), and (\ref{P-k-final}),  the Lagrangians (\ref{action-A-1})-(\ref{action-Phi-1}) can be manipulated, as first shown in \cite{Braaten:1991gm}, to the forms
\ba
\label{action-A-2}
{\cal L}^{A}_2(x) &=& C_\Pi \int \frac{d^3p}{(2\pi )^3} \,
\frac{f_\Pi ({\bf p})}{E_p} \,
F_{\mu \nu} (x) {p^\nu p^\rho \over (p \cdot \partial)^2} F_\rho^{\mu} (x) ,
\\ [2mm]
\label{action-Psi-2}
{\cal L}^{\Psi}_2(x) &=& C_\Sigma
\int \frac{d^3p}{(2\pi )^3} \, \frac{ f_\Sigma({\bf p})}{E_p} \,
\bar{\Psi}(x) {p \cdot \gamma \over p\cdot \partial} \Psi(x) ,
\\ [2mm]
\label{action-Phi-2}
{\cal L}^{\Phi}_2(x) &=& - C_P
\int \frac{d^3p}{(2\pi )^3} \, \frac{f_P({\bf p})}{E_p} \;
\Phi^*(x) \Phi(x) ,
\ea
where the operator inverse to ${p \cdot \partial}$ acts as
\be
\frac{1}{p \cdot \partial} \Psi(x) \equiv i\int \frac{d^4k}{(2\pi)^4} \frac{e^{i k \cdot x}}{p \cdot k} \Psi(k).
\ee
The operator $({p \cdot \partial})^{-2}$ is defined analogously. 

The $n-$point functions with $n > 2$, which are generated by the actions (\ref{action-A-2})-(\ref{action-Phi-2}), identically vanish, as the actions are quadratic in fields. We also observe that the action of scalars (\ref{action-Phi-2}) is gauge invariant for every theory which includes the scalar field. Moreover, the gauge boson action (\ref{action-A-2}) is invariant as well but only in the Abelian theories.  The fermion action is gauge dependent in all theories under consideration. Therefore, the fermion action and, in general, the gauge boson action need to be modified to comply with the principle of gauge invariance. This is achieved by simply replacing the usual derivative $\partial^\mu$ by the covariant derivative $D^\mu$ in Eqs.~(\ref{action-A-2}) and (\ref{action-Psi-2}). Thus, we obtain
\ba
\label{action-A-HL}
{\cal L}^{A}_{\rm HL}(x) &=& C_\Pi \int \frac{d^3p}{(2\pi )^3} \,
\frac{f_\Pi({\bf p})}{E_p} \,
F_{\mu \nu} (x) {p^\nu p^\rho \over (p \cdot D)^2} F_\rho^{\mu} (x) ,
\\ [2mm]
\label{action-Psi-HL}
{\cal L}^{\Psi}_{\rm HL}(x) &=& C_\Sigma
\int \frac{d^3p}{(2\pi )^3} \, \frac{ f_\Sigma({\bf p})}{E_p} \,
\bar{\Psi}(x) {p \cdot \gamma \over p\cdot D} \Psi(x) ,
\\ [2mm]
\label{action-Phi-HL}
{\cal L}^{\Phi}_{\rm HL}(x) &=& - C_P
\int \frac{d^3p}{(2\pi )^3} \, \frac{f_P({\bf p})}{E_p} \;
\Phi^*(x) \Phi(x) .
\ea
The forms of covariant derivatives present in Eqs.~(\ref{action-A-HL}) and (\ref{action-Psi-HL}) depend on the theory under consideration. In the electromagnetic theories, the derivative in the gauge boson action (\ref{action-A-HL}) is, as already mentioned,  the usual derivative while that in the fermion action  (\ref{action-Psi-HL}) is $D^\mu = \partial^\mu - ie A^\mu$. The operator $(p \cdot D)^{-1}$ acts as 
\be
\label{inv-cov}
\frac{1}{p \cdot D} \Psi(x) \equiv \frac{1}{p \cdot \partial} \sum_{n=0}^\infty
\Big(- ie p \cdot A(x) \frac{1}{p \cdot \partial}\Big)^n  \Psi(x).
\ee
In the ${\cal N}=4$ super Yang-Mills the covariant derivatives in Eqs.~(\ref{action-A-HL}) and (\ref{action-Psi-HL})  are both in the adjoint representation of ${\rm SU}(N_c)$ gauge group. The formula (\ref{inv-cov}) should be then appropriately modified. In QCD, the covariant derivative in Eq.~(\ref{action-A-HL}) is in the adjoint representation but that in Eq.~(\ref{action-Psi-HL}) is in the fundamental one. As already mentioned,  there is an extra factor 1/2 in the r.h.s of Eq.~(\ref{action-Phi-HL}) in case of ${\cal N}=4$ super Yang-Mills.

The hard-loop actions (\ref{action-A-HL}), (\ref{action-Psi-HL}), and (\ref{action-Phi-HL}) are all of the universal form for a whole class of gauge theories. However, the case of Abelian fields differs from that of nonAbelian ones. In the electromagnetic theories the gauge boson and scalar actions are quadratic in fields.  Therefore, the $n-$point functions generated by these actions vanish for $n>2$. Only the fermion action generates the non-trivial three-point and higher functions. The action (\ref{action-Psi-HL}) is, in particular, responsible for a modification of the electromagnetic vertex. In the nonAbelian theories, both the gauge boson and fermion actions generate  the non-trivial three-point and higher functions. Therefore, the gluon-fermion, three-gluon, and four-gluon couplings are all modified.

\section{Discussion}
\label{sec-discussion}

We have shown that the hard-loop self-energies of gauge, fermion, and scalar fields are of the universal structures and so are the effective actions of QED, scalar QED, $\mathcal{N}=1$ super QED, Yang-Mills, QCD, and $\mathcal{N}=4$ super Yang-Mills. One asks why the universality occurs physically. Taking into account a diversity of the theories - various field content and microscopic interactions - the uniqueness of the hard-loop effective action is rather surprising. 

To better understand the problem in physical terms, let us consider the QED plasma of spin 1/2 electrons and positrons and the scalar QED plasma of spin 0 particles and antiparticles. The universality of hard-loop action means that neither effects of quantum statistics of plasma constituents are observable nor the differences in elementary interactions which govern the dynamics of the two systems. Both facts can be understood as follows. The hard-loop approximation requires that the momentum at which a plasma is probed, that is the wavevector $k$, is much smaller than the typical momentum of a plasma constituent $p$. Therefore, the length scale, at which the plasma is probed, $1/k$, is much greater than the characteristic de Broglie wavelength of plasma particle, $1/p$. The hard-loop approximation thus corresponds to the classical limit where fermions and bosons of the same masses and charges are not distinguishable. The fact that the differences in elementary interactions are not seen results from the very nature of gauge theories - the gauge symmetry fully controls the interaction. And the hard-loop effective actions obey the gauge symmetry. 

The universality of hard-loop actions has far-reaching physical consequences: the characteristics of all plasma systems under consideration, which occur at the soft scale, are qualitatively the same. In particular, spectra of collective excitations of gauge, fermion, and scalar fields are the same. Therefore, if the electromagnetic plasma with a given momentum distribution is, say, unstable, the quark-gluon plasma with this momentum distribution is unstable as well. We conclude that in spite of all differences, the plasma systems under consideration are very similar to each other at the soft scale. However, the hard-loop approach breaks down for the momenta at and below the magnetic sale. Then, systems governed by different theories can behave very differently. In particular, the QED plasma is very different from the QCD one, as in the latter case effects of confinement apparently appear at the magnetic scale. Recently, there have been undertaken several efforts to extend methods of the hard-loop approach to the ultrasoft scale \cite{Maas:2011se,Gao:2014rqa,Hidaka:2011rz,Blaizot:2014hka,Su:2014rma}. These efforts explicitly show limitations of the universality we have elaborated on here.

\section*{Acknowledgments}

This work was partially supported by the Polish National Science Centre under Grant No. 2011/03/B/ST2/00110.


\end{document}